# The initiation temperatures in nanothermite reactions

V.G. Myagkov

Kirensky Institute of Physics, Federal Research Center KSC SB RAS, 660036 Krasnoyarsk, Russia

ORCID   0000-0002-3276-0309

E-mail: miagkov@iph.krasn.ru

**Abstract**

An assumption on general regularities and chemical mechanisms of solid-state reactions in nanofilms and nanothermite mixtures is made. It is demonstrated that the moving force of all Al-based nanothermite reactions is the synthesis of the $Al_2O_3$ phase with a high negative enthalpy of formation. According to this assumption, all Al-based nanothermite reactions should have the same initiation temperature $T_{in}$. Indeed, an analysis of literature data shows that the synthesis of the $Al_2O_3$ phase in $Al/Fe_2O_3$, $Al/CuO$, $Al/Co_3O_4$, $Al/MoO_3$, $Al/Bi_2O_3$, and $Al/NiO$ nanothermite mixtures starts at the same temperature $T_{in} \sim 510°C$. We also demonstrate the same initiation temperatures ~250°C, ~300°C, and ~180°C for Zr-, Mg- and In-based nanothermite reactions, respectively. It is predicted that nanothermite reactions based on other fuels have their own initiation temperatures.

**Keywords:** Solid-state reactions, Nanofilms, Nanothermite reactions, First phase, Initiation temperature.

## 1. Introduction

The control and predictability of the synthesis of new materials is one of the most fundamental challenges in material science. The traditional approach to this problem is to use the prediction of crystal structure from first principles [1- 4]. However, now the lack of control and predictability are indeed notorious characteristics of the synthesis of new materials [5]. The prediction of the phases in binary systems that will be formed during the thin film solid-state reaction has been a subject of numerous studies, and different empirical rules have been developed for predicting first phase formation [6-14].

Studies of solid-state reactions in nanolayers have shown three fundamental features that strongly distinguish them from bulk powders:

(i) Formation of only the first phase at the film reagent interface at a certain temperature $T_{in}$ called the initiation (formation) temperature $T_{in}$. As the annealing temperature is increased, other phases can occur and form the phase sequence [11-14].

(ii) The threshold of the reaction, characterized by intense intermixing at the interface and formation of compounds, arises at temperatures above the initiation (formation) temperature $T_{in}$. The values of $T_{in}$ in the first phase can be about room temperature [15, 16] or even at cryogenic temperatures [17-20]

(iii) Migration of the dominant diffusing species through the interface during first phase formation [13, 21, 22].

The formation of only the first phase among equilibrium phases, low initiation temperatures, and migration of the dominant diffusing species are unique, unexplained features of solid-state reactions in nanofilms. From the above results it follows that the first phase and its initiation temperature $T_{in}$ are control characteristics of the thin film solid-state reactions.

## 2. Presentation of the hypothesis

Over the past decade, investigations focused on the nanoenergetic materials, such as reactive multilayer thin films [23, 24] and nanostructured reactive mixtures [25-27]. Thermite mixtures belong to a wide class of energetic materials that comprise a metal fuel (e.g., Al, Mg, or B) and an oxidizer (e.g., $Fe_2O_3$, $MoO_3$, CuO, $Bi_2O_3$, or $WO_3$). These mixtures react with a lot of heat release; therefore, the thermite reactions often occur in a self-sustaining mode [28-30]. Typical thermite mixtures contain micron particles and have a combustion wave velocity of 1-20 m/s. In recent years, there has been an increasing interest in nanothermites (superthermites) where the particle size is reduced to a few nanometers. In nanothermite mixtures, the combustion wave velocity grows to 1000 m/s [25-27]. Studies of nanothermite combustion are mainly aimed at measuring the combustion wave velocity, burning temperature, delay time, and their dependences on density, morphology, and composition of the reaction mixture. Despite the intense investigations of thermite reactions, their general regularities and mechanisms remain unclear. Currently, the classical nanothermite Goldschmidt reaction $Fe_2O_3 + 2Al = Al_2O_3 + 2Fe$ and other Al-based reactions are well studied.

In this work, we extend the existing concepts of the first phase and its initiation temperature $T_{in}$, which describe the initial stage of solid-state reactions in nanofilms onto thermite reactions and demonstrate that in all Al-based nanothermite mixtures the synthesis of the $Al_2O_3$ phase starts at the same initiation temperature $T_{in} \sim \mathbf{510°C}$. These results open up a way for understanding the exclusive role of the initiation temperature $T_{in}$ in the solid-state reactions at the nanoscale. Our previous studies and the analysis of solid-state thin film reactions for many bilayers have shown that the initiation temperatures $T_{in}$ often coincide with temperatures $T_K$ of structural phase transformations ($T_{in} = T_K$). In particular, initiation temperatures $T_{in}$(Cu/Au) and $T_{in}$(Ni/Fe) of reactions in the Cu/Au and Ni/Fe bilayers coincide with the minimum temperature of the order-disorder phase transition in Cu-Au [31] and the eutectoid decomposition temperature in the Fe-Ni system [32], respectively. In Ni/Al and Cd/Au films, the reactions start at the temperature of the inverse martensitic transformation in the Ni-Al [33] and Cd-Au [34] binary systems, respectively. The equality $T_{in} = T_K$ was also established for the eutectic reactions, superionic transition, and spinodal decomposition in Al/Ge [35], Se/Cu [36], and Mn/Ge [37, 38] films, respectively. The equality $T_{in} = T_K$ indicates the common nature of chemical interactions that control both solid-state reactions in thin films and solid-state transformations.

## 3. Testing of the hypothesis

The above results demonstrate that the first phase and its temperature $T_{in}$ are the fundamental characteristics of a reaction bilayer. It follows that low-temperature reactions occur only between the reacting layers whose reaction products have low-temperature solid-state transformations. Therefore, the study of phase sequences in reaction couples makes it possible to refine and supplement the phase equilibrium diagrams especially in low-temperature part. Thus, recently we investigated the reaction in Ge/Mn bilayers and confirmed the existence of spinodal decomposition in an in Ge-Mn system at 120°C. It is important to note that the formation of the first $Mn_5Ge_3$ phase was independent of whether Mn and Ge atoms are in a solid solution or in Ge/Mn bilayers [37, 38].

The one important characteristic of reactive multilayer films is the ignition temperature $T_{ig}$, which can be defined as a minimum temperature of onset of a self-sustaining reaction for a given experiment [23, 24]. As known, the self-sustaining regime of reaction arises when the rate of heat generation $Q_{reaction}$ overcomes the rate of heat losses $Q_{loss}$ ($Q_{reaction} > Q_{loss}$). Unlike $T_{ig}$, the initiation temperature $T_{in}$ is the start temperature of reactions at which the rate of heat generation $Q_{reaction}$ is less than the rate of heat losses $Q_{loss}$ ($Q_{reaction} < Q_{loss}$) and so the initiation temperature $T_{in}$ is always less than the ignition temperature $T_{ig}$ ($T_{in} < T_{ig}$). As discussed above the initiation temperature $T_{in}$ is the threshold temperature: no reaction below $T_{in}$ and reaction initiate just the temperature of sample overcomes $T_{in}$. Thus, the initiation temperature $T_{in}$ is the characteristic temperature of our given reaction couple. In contrast to $T_{in}$, the ignition temperature $T_{ig}$ is a kinetic quantity that depends on the heating rate and the rate of heat loss. Nevertheless, Frits et al. have recently shown that, similar to $T_{in}$, the ignition temperature $T_{ig}$ in Ni/Al multilayers with a given bilayer thickness is a threshold temperature because in hot plate experiments the multilayers do not ignite when the specimens are heated to temperatures just 1°C below $T_{ig}$ [39].

The enthalpy of formation of the first phase is a good measure of the free energy variation during the solid-state interaction; therefore, the heats of formation were used in several initial models to predict the first phase and phase sequence formation. Pretorius *et* al. [13] proposed an effective model for the enthalpy of formation, which was successfully used for predicting the first phase formation in many binary systems.

As mentioned above, with the increasing of the temperature of the bilayer above $T_{in}$ leads to the beginning of intermixing of the reagents and first phase synthesis on the interface and consequently physical characteristics of the film samples, such as electrical resistance, magnetization, transparence, and heat release, begin to radically change. Obviously, the start temperature of these changes is the reaction initiation temperature $T_{in}$. In most cases the energetic properties of thermite nanocomposites were investigated by differential thermal analysis (DTA), thermogravimetric analysis (TGA), and differential scanning calorimetry (DSC). In this case the initiation temperature $T_{in}$ is the temperature at which heat release starts. An important characteristic of the DSC curves is also the exothermic peak temperature, which, unlike the initiation temperature $T_{in}$, depends on the heat removal conditions from the reaction zone.

It is important to note that the contaminants that form on the reagent interface during various methods of sample preparation (especially for chemically produced samples) can form thin barrier layers that slightly change the initiation temperature of $T_{in}$ but do not suppress the reaction. An error in finding the exact value of the initiation temperature $T_{in}$ can also follow from the certain inaccuracy in determining $T_{in}$ from DTA, TGA and DSC plots. To find the exact $T_{in}$ value, low heating rates are required. Therefore, we referred only to the studies in which the heat release curves were obtained at minimum heating rates (5,10 or 20 °C/min).

The main results of the work are summarized in the schematic diagram in Figure 1, showing the initiation temperature $T_{in}$ ~ 510°C of the $Al_2O_3$ phase in $Al/Fe_2O_3$, $Al/Co_3O_4$, $Al/NiO$, $Al/MnO_2$, $Al/Bi_2O_3$, $Al/CuO$, $Al/MoO_3$ nanothermite reactions and oxidation of Al nanomaterials.

## 3. Implication of the hypothesis

### 3.1. The initiation temperature $T_{in}$ ~ 510°C of the Al-based nanothermite reactions

It can be seen from Tables 1 – 4 that, taking into account the errors in measuring the DTA, TGA, DSC curves, and magnetic and resistivity plots by different authors, the initiation temperatures of all Al-based nanothermite reactions are centered around ~ **510** °C. As can be seen from Tables 1- 4, at heating rates 5, 10, 20 °C/min the heat release curves also have closely exothermic peak temperatures. It must be noted that at a heating rate of 5 °C/min, the oxidation of nanosized aluminum powders also starts at about **510 °C** [60, 92-100]. Recently it was shown that the DSC curves of the $NiCo_2O_4$/Al core-shell nanowires thermite film [101] and three-dimensional ordered macroporous $NiFe_2O_4$/Al nanothermites [102] have an exothermic peak with an initiation temperature ~ 530°C and an exothermic peak temperature ~ 600°C. This unambiguously proves, that the synthesis of $Al_2O_3$ is the driving force of the thermite reaction $8Al + 3Ni(Co\ or\ Fe)_2O_4 \rightarrow 4Al_2O_3 + 3Ni + 6(Co\ or\ Fe)$ in the $NiCo_2O_4$/Al and $NiFe_2O_4$/Al nanocomposites.

These results suggest the following reaction mechanism: below the initiation temperature $T_{in} <$ ~ **510** °C, Al and O atoms remain chemically neutral. At $T_{in} >$ ~ **510** °C, strong chemical interactions occur between Al and O atoms that break old chemical bonds causing the directed atomic migration to the reaction zone and the synthesis of $Al_2O_3$ regardless of the system they exist in. Therefore, the initiation temperature $T_{in}$ ~ **510** °C is a universal parameter of all Al-based nanothermite reactions.

### 3.2. The initiation temperature $T_{in}$ ~ 250 °C of the Zr-based nanothermite reactions

The driving force of the Zr-based nanothermite reactions is the formation of the $ZrO_2$ phase, which has a relatively high negative enthalpy of formation ($\Delta H_f$ = -601 kJ/mol). The analysis of the synthesis of Co-$ZrO_2$ and Fe-$ZrO_2$ ferromagnetic nanocomposites show, that thermite reactions in the Zr/$Co_3O_4$ [103] and Zr/$Fe_2O_3$ [104] thin films have the same initiation temperature $T_{in}$ (Zr/$Co_3O_4$) = $T_{in}$(Zr/$Fe_2O_3$) ~ 250 °C. This value $T_{in}$ ~ 250 °C is in agreement with the initiation temperature $T_{in}$(Zr/CuO) ~ 250 °C in the Zr/CuO nanothermite mixture [105] and with the study where the $ZrO_2$ nanoparticles were obtained by glycothermal processing [106]. From these facts we predict the same initiation temperature $T_{in}$ ~ 250 °C for all the Zr-based nanothermite reactions.

### 3.3. The initiation temperature $T_{in}$ ~ 450 °C of the Mg-based nanothermite reactions

The driving force of the Mg-based nanothermite reactions is the synthesis MgO phase having a high negative enthalpy of formation ($\Delta H_f$ = -1097 kJ/mol) with some low enthalpy of formation of $Al_2O_3$ ($\Delta H_f$ = -1676 kJ/mol). Using the results of this paper [105], we determined that the thermite multilayer Mg/CuO stacks the initiation temperature $T_{in}$(Mg/CuO) ~ 450 °C. The nearest value of the initiation temperature has nanoenergetic Mg/CuO core/shell arrays, which exhibit an onset reaction temperature (~ 450 °C) [107]. The initiation temperatures are close to ~ 450 °C in fresh CuO/Mg/fluorocarbon nanoenergetic composites [108] and μm-Mg/nmCuO thermite mixtures prepared by physical and ultrasonic mixing [109]. Magnesium particles with a nominal size of 6 μm begin an intensive oxidation above 500 °C, which suggests that the starting temperature of the oxidation of nanosized magnesium powders is below 500 °C [110]. In these papers the initiation temperature $T_{in}$ was defined as the temperature at which heat release started using the curves of differential thermal analysis (DTA), thermogravimetric analysis (TGA), and differential scanning calorimetry (DSC). Although the exact value of the initiation temperature of Mg-based nanothermite reactions remains unknown, it lies around 450 °C.

### 3.4. The initiation temperature $T_{in}$ ~ 180 °C of the In-based nanothermite reactions

Recently, we have demonstrated a new way to synthesize ferromagnetic Fe-$In_2O_3$ and Co-$In_2O_3$ nanocomposite thin films using the new thermite reactions $Fe_2O_3$ + 2In = $In_2O_3$ + 2Fe [111] and $3Co_3O_4$ + 8In + $4In_2O_3$ + 9Co [112] which starts above the initiation temperature $T_{in}$ = 180 °C - 190 °C with the predominant formation of the Fe, Co and $In_2O_3$ phases. In [113] the general strategy of the fabrication of metal oxide films has been proposed. The choice of suitable combustion precursors, containing indium nitrates as oxidizers and urea or acetyl acetone as fuel, were used for the low-temperature fabrication of $In_2O_3$ films. The initiation temperatures of combustion reactions, measured from DTA and DSC curves, are in the range 175 °C - 200 °C [113-115]. Our studies of the self-sustaining oxidation of In films have shown that the initiation temperature is ~ 180 °C [116]. The emergence of a strong chemical interaction between the In and O atoms above 180 °C may be the cause of rapid crystallization of amorphous $In_2O_3$ films within the temperature range 165-210 °C [117, 118]. The above data suggests that for the In-based nanothermites, oxidation of In and temperature crystallization of amorphous $In_2O_3$ films should have the same initiation temperature $T_{in}$ ~ 180 °C.

It is worth noticing that all nanothermite reactions have low initiation temperatures and occur in the solid state (excluding In-based reactions). On the conceptual level, the initiation temperatures are ignored in modern models described the thin-film solid-state reactions. In contrast to this, our approach assumes the decisive role of the leading (first) phase and its initiation temperature not only in thermite reactions but also in other types of chemical reactions.

Progress in understanding and predicting the structural transformations and reactions in the solid state is limited by a lack-of-knowledge inference about the chemical interactions at nanoscale. Undoubtedly, future investigations of thin-film solid-state reactions and nanothermite reactions will discover new amazing properties of chemical interactions in solids.

## 4. Conclusions

The main concepts of this study are the first phase and its initiation temperature $T_{in}$, which describe thin-film solid-state reactions and were extended onto nanothermite reactions. The paper results prove that all Al-based nanothermite reactions have the same initiation temperature. The analysis of the initiation temperatures reported in the literature and our data has shown that the synthesis of the $Al_2O_3$ phase in $Al/Fe_2O_3$, $Al/CuO$, $Al/Co_3O_4$, $Al/MoO_3$, $Al/Bi_2O_3$, $Al/NiO$ and $Al/Mn_2O_3$ nanothermite mixtures starts at a temperature of about **510** °C. The same initiation temperatures ~250 °C, ~450 °C, and ~180 °C also have Zr-, Mg- and In-based nanothermite reactions, respectively. Finally, these findings predict that nanothermite reactions based on other fuels (e.g., Ti and B) must have their own initiation temperatures. This approach can be widely applicable in the study of the multicomponent thin-film solid-state reactions.

**Acknowledgements**
I thank Zhigalov VS, Bykova LE and Matsynin AA for fruitful discussions and advice.

**Compliance with ethical standards**
Conflict of interest
The author has no conflict of interest
related to this paper.

**Figure 1**. The schematic illustration of the initiation temperature $T_{in} \sim 510\ °C$ for the Al-based nanothermite reactions with $Fe_2O_3$, $Co_3O_4$, $NiO$, $MnO_2$, $Bi_2O_3$, $CuO$, $MoO_3$ oxidizers and the oxidation of Al nanomaterials. The initiation temperature $T_{in} \sim 510\ °C$ is characteristic of the leading $Al_2O_3$ phase, which has a high negative enthalpy of formation ($\Delta H_f = -1676\ kJ/mol$) and is the driving force of all the Al-based nanothermite reactions.

**Table 1.** Summary of Al-based nanothermite reactions with $Fe_2O_3$ oxidizer and with an associated record of the initiation temperatures.

**Table 2.** Summary of Al-based nanothermite reactions with $CuO$ oxidizer and with an associated record of the initiation temperatures.

**Table 3.** Summary of Al-based nanothermite reactions with $Bi_2O_3$, $MoO_3$ oxidizers and with an associated record of the initiation temperatures.

**Table 4.** Summary of Al-based nanothermite reactions with $Co_3O_4$, $NiO$, $MnO_2$, oxidizers and with an associated record of the initiation temperatures.

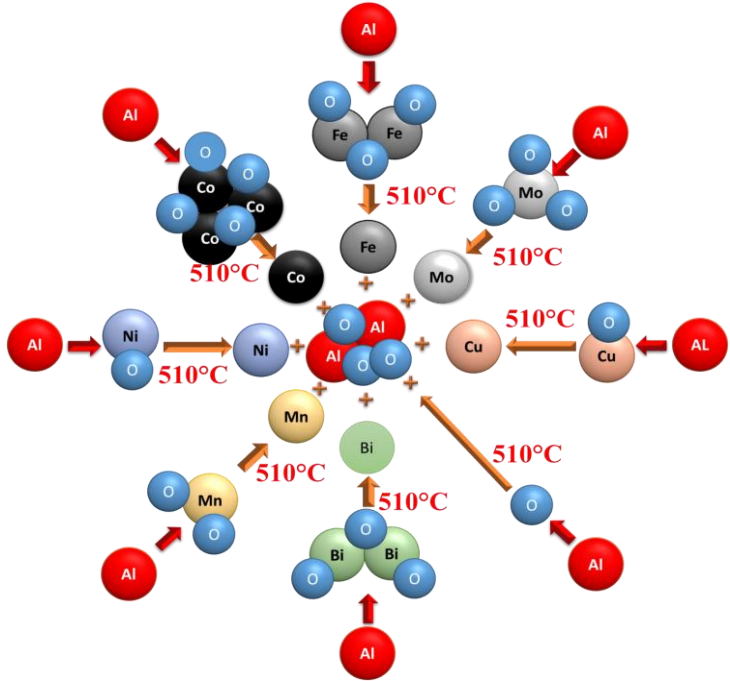

**Figure 1**. The schematic illustration of the initiation temperature $T_{in} \sim 510$ °C for the Al-based nanothermite reactions with $Fe_2O_3$, $Co_3O_4$, NiO, $MnO_2$, $Bi_2O_3$, CuO, $MoO_3$ oxidizers and the oxidation of Al nanomaterials. The initiation temperature $T_{in} \sim 510$ °C is characteristic of the leading $Al_2O_3$ phase, which has a high negative enthalpy of formation ($\Delta H_f = -1676$ kJ/mol) and is the driving force of all the Al-based nanothermite reactions.

Table 1. Summary of Al-based nanothermite reactions with $Fe_2O_3$ oxidizer and with an associated record of the initiation temperatures.

| Nanothermite Systems | Initiation Temperature $T_{in}$ (°C) | Techniques | Temperature Rate $\eta$ (K/min.) | Exothermic Peak Temperature (°C) | Samples | Refs. |
|---|---|---|---|---|---|---|
| Al/$Fe_2O_3$ | ~ 490 | DSC | 10 | ~ 530 | $Fe_2O_3$/Al xerogel nanocomposites | [40] |
| Al/$Fe_2O_3$ | ~ 530 | TGA/DSC | 10 | ~ 550 | $Fe_2O_3$/Al xerogel nanocomposites | [41] |
| Al/$Fe_2O_3$ | ~ 510 | DSC | 10 | 588 | $Fe_2O_3$ nanotubes, Al nanocomposites | [42] |
| Al/$Fe_2O_3$ | ~ 520 | DSC | 10 | ~ 545 | $Fe_2O_3$ nanoporous particles, Al powder | [43] |
| Al/$Fe_2O_3$ | ~ 520 | DSC | 20 | 548 | $Fe_2O_3$/Al nanothermite membranes. | [44] |
| Al/$Fe_2O_3$ | ~ 530 | Magnetization vs temperature | > 10 | | $Fe_2O_3$/Al bilayers. | [45] |
| Al/$Fe_2O_3$ | ~ 525 | TGA/DSC | 8-30 | ~ 590 | $Fe_2O_3$/Al nanothermite membranes | [46] |
| Al/$Fe_2O_3$ | 507-508 | TGA/DSC | 10 | ~ 590 | $Fe_2O_3$, Al nanoparticles | [47] |
| Al/$Fe_2O_3$ | ~ 520 | TGA/DTA | 20 | ~ 620 | Al/$Fe_2O_3$ nanothermite film | [48] |
| Al/$Fe_2O_3$ | ~ 490 | TGA/DTA | 10 | ~ 570 | Porous Core/Shell Structure $Fe_2O_3$/Al | [49] |
| Al/$Fe_2O_3$ | ~ 490 | DSC | 10 | ~ 570 | core–shell structured Al@$Fe_2O_3$ nanothermite | [50] |
| Al/$Fe_2O_3$ | ~ 500 | DSC | 20 | ~ 550 | Al@$Fe_2O_3$ nanocomposites | [51] |
| Al/$Fe_2O_3$ | ~ 500 | TGA/DSC | 10 | ~ 590 | Al/$Fe_2O_3$/MWCNT nanostructured energetic materials | [52] |
| Al/$Fe_2O_3$ | ~ 480 | DSC | 20 | ~ 560 | nanostructured energetic materials sol–gel–Al/$Fe_2O_3$ | [53] |
| Al/$Fe_2O_3$ | 504-525 | DSC/TG | 5 | 546-569 | n-Al/n-$Fe_2O_3$ nanothermite | [54] |

Table 2. Summary of Al-based nanothermite reactions with CuO oxidizer and with an associated record of the initiation temperatures.

| Nanothermite Systems | Initiation Temperature $T_{in}$ (°C) | Techniques | Temperature Rate $\eta$ (K/min.) | Exothermic Peak Temperature (°C) | Samples | Refs. |
|---|---|---|---|---|---|---|
| Al/CuO | ~ 520 | DTA/DSC | 15/5 | 580/560 | CuO nanowires, Al films. | [55] |
| Al/CuO | 515 | DSC | 10 | ~ 530 | Al/CuO bilayes | [56] |
| Al/CuO | ~ 520 | DTA/DSC | 5 | ~ 560 | Al/CuO multilayers | [57] |
| Al/CuO | ~ 500 | TG/DSC | 15 | ~ 540 | CuO nanowires coated with deposited nano-Al | [58] |
| Al/CuO | ~ 520 | DTA/DSC | 10 | ~ 566 | CuO nanowires, Al nanoparticles | [59] |
| Al/CuO | ~ 500 | TGA/DSC | 10 | ~ 560 | Al, CuO nanoparticles | [60] |
| Al/CuO | ~ 500 | DSC/TG | 10 | ~ 620 | nanoAl, CuO nano-array | [61] |
| Al/CuO | ~ 500 | DSC | 10 | ~ 550 | CuO/Al multilayers | [62] |
| Al/CuO | ~ 520 | DSC | 5 | ~ 540 | Al, CuO nanoparticles | [63] |
| Al/CuO | ~ 500 | DSC | 10 | ~ 560 | Al, CuO nanoparticles | [64] |
| Al/CuO | ~ 520 | DSC | 5 | ~ 560 | Al/CuO nanoparticles core-shell structures | [65] |
| Al/CuO | ~ 530 | DG/DSC | 10 | ~ 590 | Al, CuO nanoparticles | [66] |
| Al/CuO | ~ 480 | DSC | 10 | ~ 550 | Al, CuO nanoparticles | [67] |
| Al/CuO | ~ 480 | DSC | 10 | ~ 550 | Al/CuO core/shell arrays | [68] |

Table 3. Summary of Al-based nanothermite reactions with $Bi_2O_3$, $MoO_3$ oxidizers and with an associated record of the initiation temperatures.

| Nanothermite Systems | Initiation Temperature $T_{in}$ (°C) | Techniques | Temperature Rate η (K/min.) | Exothermic Peak Temperature (°C) | Samples | Refs. |
|---|---|---|---|---|---|---|
| Al/MoO$_3$ | 476 | DTS | 10 | ~ 500 | MoO$_3$ nanoparticles, Al micro particles | [69] |
| Al/MoO$_3$ | ~ 500 | DTS | 10 | ~ 550 | Al/MoO$_3$ nanocomposites | [70] |
| Al/MoO$_3$ | ~ 520 | DSC | 20 | ~ 560-590 | Reactive multilayer films | [71] |
| Al/MoO$_3$ | ~ 475-515 | DTA/DSC | | ~ 550 | nano-Al, MoO$_3$ nanobelts | [72] |
| Al/MoO$_3$ | ~ 520 | TG/DSC | 10 | ~ 560 | Al/MoO$_3$ xerogel nanocomposite | [73] |
| Al/MoO$_3$ | ~ 440 | TGA/DSC | 20 | ~ 520 | (2D) molybdenum trioxide, Al nanoparticles | [74] |
| Al/MoO$_3$ | ~ 480 | DSC | 10 | ~ 550 | Al, MoO$_3$ nanoparticles | [67] |
| | | | | | | |
| Al/Bi$_2$O$_3$ | ~ 520 | DTS | 10 | ~ 550 | Al, Bi$_2$O$_3$ nanopowders | [75] |
| Al/Bi$_2$O$_3$ | ~ 510 | TG/DTS | 10 | ~ 572/589 | Al, Bi$_2$O$_3$ nanofilms | [76] |
| Al/Bi$_2$O$_3$ | ~ 480 | DSC | 20 | 591 | Al, Bi$_2$O$_3$ nanoparticles | [77] |
| Al/Bi(OH)$_3$ | ~ 520 | DSC | 20 | 603 | Al-Bi(OH)$_3$ nano-thermite | [78] |
| Al/Bi$_2$O$_3$ | ~ 480 | DSC | 10 | ~ 550 | Al, Bi$_2$O$_3$ nanoparticles | [67] |

Table 4. Summary of Al-based nanothermite reactions with $Co_3O_4$, NiO, $MnO_2$, oxidizers and with an associated record of the initiation temperatures.

| Nanothermite Systems | Initiation Temperature $T_{in}$ (°C) | Techniques | Temperature Rate $\eta$ (K/min.) | Exothermic Peak Temperature (°C) | Samples | Refs. |
|---|---|---|---|---|---|---|
| Al/$Co_3O_4$ | ~ 520 | DTA/TGA | 10 | 574/569 | $Co_3O_4$ nanowires, nano-Al | [79] |
| Al/$Co_3O_4$ | ~ 551 | DSC | 10 | ~ 600 | $Co_3O_4$/Al core/shell nanowires | [80] |
| Al/$Co_3O_4$ | ~ 500 | DSC | 10-30 | ~ 560-590 | Al/$Co_3O_4$ nanothermites film | [81] |
| Al/$Co_3O_4$ | ~ 500 | DSC | 10 | ~ 560-590 | the $Co_3O_4$ particles were embedded in the aluminum particles | [82] |
| Al/$Co_3O_4$ | ~ 535 | TG-DSC | 10 | ~ 605 | $Co_3O_4$/nanoAl | [83] |
| Al/$Co_3O_4$ | ~ 510 | Magnetization and electrical resistance as a function temperature | ~ 4 | | $Co_3O_4$/Al nanofilms | [84-85] |
| Al/NiO | ~ 475-515 | DTA/DSC | 10 | ~ 550 | NiO nanowires, Al nanopaticles | [86] |
| Al/NiO | ~ 490 | DSC | 10 | ~ 549 | NiO nanowires, Al nanoparticles | [87] |
| Al/NiO | ~ 460 | DSC | 20 | ~ 530-565 | Three-dimensionally ordered macroporous NiO/Al nanothermite film. | [88] |
| Al/NiO | ~ 520 | DSC | 20 | ~ 590 | Nano-Al/NiO thermite films | [89] |
| Al/NiO | ~ 516 | DGA/DSC | 10 | ~ 535 | Al/NiO thermite Film | [60] |
| Al/$Mn_2O_3$ | 474-518 | DSC | 20 | 563-599 | $Mn_2O_3$ macroporous skeleton, Al films | [90] |
| $MnO_2$/$SnO_2$/n-Al | ~520 | DSC | 20 | 600 | $MnO_2$/$SnO_2$/n-Al ternary thermite membrane | [91] |